\documentclass[12pt]{article}

\usepackage{latexsym}
\usepackage{amssymb,amsfonts,amsmath}
\usepackage{graphicx} 
\usepackage{indentfirst}
\usepackage{bbm}
\usepackage{amssymb}
\usepackage{verbatim}
\usepackage{amsmath, amsthm,amssymb}
\usepackage{mathrsfs}
\usepackage{hyperref}
\usepackage{amsfonts}
\usepackage{dsfont}

\parskip=\medskipamount

\newcommand {\cD}{{\cal D}}
\newcommand {\cE}{{\cal E}}
\newcommand {\cF}{{\cal F}}

\newcommand {\cL}{{\cal L}}

\newcommand {\cN}{{\cal N}}

\newcommand {\cS}{{\cal S}}

\newcommand {\cU}{{\cal U}}
\newcommand {\cV}{{\cal V}}

\newcommand {\cY}{{\cal Y}}

\def\a{\alpha}

\def\b{\beta}
\def\c{\chi}
\def\d{\delta}

\def\g{\gamma}
\def\G{\Gamma}

\def\j{\psi}
\def\k{\kappa}

\def\m{\mu}

\def\q{\theta}
\def\r{\rho}
\def\s{\sigma}

\def\D{\Delta}
\def\F{\Phi}
\def\J{\Psi}
\def\L{\Lambda}

\def\P{\Pi}

\def\S{\Sigma}

\def\tr{{\rm tr}}
\def\rd{{\rm d}}
\def\ri{{\rm i}}
\def\re{{\rm e}}

\newcommand{\ad}{{\dot{\alpha}}}                           
\newcommand{\bd}{{\dot{\beta}}}                            
\newcommand{\cDB}{{\bar\cD}}                            
\newcommand{\DB}{\bar{D}}

\newcommand{\pa}{\partial}                           
\newcommand{\hf}{\frac12}

\newcommand{\vf}{\varphi}

\newcommand{\be}{\begin{equation}}
\newcommand{\ee}{\end{equation}}
\newcommand{\bea}{\begin{eqnarray}}
\newcommand{\eea}{\end{eqnarray}}
\newcommand{\non}{\nonumber}
\newcommand{\bm}[1]{\mbox{\boldmath$#1$}}

\def\double #1{#1{\hbox{\kern-2pt $#1$}}}

\newcommand{\qb}{{\bar{\theta}}}

\newif\ifdtup

\newcommand{\bsubeq}{\begin{subequations}}
\newcommand{\esubeq}{\end{subequations}}

\numberwithin{equation}{section}

\begin{document}
\begin{titlepage}
\begin{flushright}
December, 2017\\
Revised: February, 2018
\end{flushright}
\vspace{5mm}

\begin{center}
{\Large \bf Nilpotent  \mbox{$\bm{\cN=1}$} tensor multiplet
}\\ 
\end{center}

\begin{center}

{\bf Sergei M. Kuzenko
} \\
\vspace{5mm}

\footnotesize{
{\it Department of Physics M013, The University of Western Australia\\
35 Stirling Highway, Crawley W.A. 6009, Australia}}  
~\\
\texttt{sergei.kuzenko@uwa.edu.au }\\

\end{center}
\vspace{2mm}

\begin{abstract}
\baselineskip=14pt
We propose a nilpotent ${\cal N} =1$ tensor multiplet describing two fields, which are
the Goldstino and the axion, the latter being realised in terms of the field strength 
of a gauge two-form. This supersymmetric multiplet is formulated in terms of a deformed real linear superfield, subject to a cubic nilpotency condition. Its couplings 
to a super Yang-Mills multiplet and supergravity are presented. 
To define a nilpotent tensor multiplet in the locally supersymmetric 
case, one has to make use of either real or complex three-form supergravity theories,
which are variant realisations of the old minimal formulation for ${\cal N} =1$ 
supergravity.
\end{abstract}

\vfill

\vfill
\end{titlepage}

\newpage
\renewcommand{\thefootnote}{\arabic{footnote}}
\setcounter{footnote}{0}

\tableofcontents{}
\vspace{1cm}
\bigskip\hrule

\allowdisplaybreaks


\section{Introduction} \label{section1}
\setcounter{equation}{0}

Following the construction of inflationary models 
with nilpotent superfields \cite{ADFS,FKL,DZ,DFKS}, 
in the last two years there has been much interest
in models for de Sitter supergravity, see
\cite{BFKVP,HY,K15,KW,SvdWW,BMST,FKRR,BHKMS,KMcAT-M,BK17}
and references therein.\footnote{The terminology ``de Sitter supergravity''
was coined in \cite{BFKVP}.} They are off-shell models for 
spontaneously broken local supersymmetry  
obtained by coupling $\cN=1$ supergravity to 
various nilpotent Goldstino superfields. 
 One of the reasons for the interest in such theories  is that 
 a positive contribution to the cosmological constant is generated once the local supersymmetry becomes spontaneously broken. 
 For instance, if the supergravity multiplet is coupled 
 to an irreducible Goldstino superfield\footnote{The 
notion of irreducible and reducible Goldstino superfields was introduced
in \cite{BHKMS}.} 
 \cite{LR79,SW,KTyler,K15,BHKMS} (with the Volkov-Akulov Goldstino \cite{VA,AV} 
being the only independent component field of the superfield), 
a {\it universal} positive contribution to the cosmological constant is generated,
which is proportional to $f^2$, with 
the parameter $f$ setting the scale of supersymmetry breaking. 
The same positive contribution is generated by the reducible Goldstino 
superfields used in the models studied in \cite{BFKVP,HY,KMcAT-M}.
There is one special reducible Goldstino superfield,
the nilpotent three-form multiplet introduced in \cite{FKRR,BK17}, which yields 
a dynamical contribution to the cosmological constant. 

Historically, 
the first off-shell model for de Sitter supergravity
was constructed  by Lindstr\"om and Ro\v{ce}k in 1979 \cite{LR79}.
They made use of the {\it irreducible} nilpotent chiral Goldstino superfield
proposed by Ro\v{c}ek \cite{Rocek}. As shown in \cite{BHKMS,KMcAT-M}, 
on the mass shell this model 
is equivalent to the one advocated in  \cite{BFKVP,HY}, which made use of the
{\it reducible}  nilpotent chiral Goldstino superfield
proposed in \cite{Casalbuoni,KS}.

Since we live in a universe dominated by dark energy and dark matter, 
and since dark energy can be sourced by a small positive 
cosmological constant, it is remarkable that spontaneously broken local 
supersymmetry provides a mechanism to generate 
a  universal positive contribution to the cosmological constant, 
which is associated with the Goldstino.
As concerns the known  candidates for dark matter
(see, e.g., \cite{Nath} for a review), the axion field is 
among the most interesting ones.  
It is natural to wonder whether there exists a constrained supermultiplet 
containing both the Goldstino and the axion. 
Such a supermultiplet  is proposed in this paper.

In the case of $\cN=1$ supersymmetry, every known scalar Goldstino superfield $X$, irreducible or reducible, obeys the quadratic nilpotency condition
\bea
X^2 =0~.
\label{0.0}
\eea
 These Goldstino superfields include:
(i) the irreducible chiral scalar proposed in  \cite{Rocek,IK78}; 
(ii) the reducible chiral scalar of  \cite{Casalbuoni,KS}; 
(iii) the deformed complex linear scalar introduced in  \cite{KTyler}; 
(iv) the complex linear scalar of  \cite{Tyler,Farakos:2015vba};
(v) the irreducible real scalar proposed in \cite{BHKMS}; 
(vi) the reducible real scalar of  \cite{KMcAT-M}. 
In the (v) and (vi) cases, 
there are actually three nilpotency conditions \cite{BHKMS,KMcAT-M}:
\begin{subequations} 
\label{0.1}
\bea
V^2&=&0~,  \label{0.1a}\\
V D_A D_B V &=&0~, 
\label{0.1b}
\\
V D_A D_B D_C V &=&0~, 
\label{0.1c}
\eea
\end{subequations}
where $D_A =(\pa_a , D_\a, \bar D^\ad)$ are the covariant derivatives of 
Minkowski superspace ${\mathbb M}^{4|4}$.\footnote{A real scalar Goldstino superfield
was briefly discussed  in  Ref. \cite{LR79} and later reviewed in \cite{SW}. However, 
only the constraint \eqref{0.1a} was explicitly given in these publications.}

The standard linear multiplet \cite{FWZ}, which is described by a real scalar 
superfield $G=\bar G$ 
constrained by $\bar D^2 G=0$, cannot be subject to any nilpotency condition 
in order to describe a Goldstino superfield. 
The point is that the $\cN=1$ tensor multiplet 
\cite{Siegel}, for which $G$ originates as  the gauge invariant field strength, has no auxiliary field.
Therefore, the constraint $\bar D^2 G=0$ has to be deformed if we wish 
to use a linear-type superfield to embed the Goldstino into.
In order to get a feeling for a suitable deformation, 
let us consider the simplest model for spontaneously broken supersymmetry,
realised in terms of a chiral scalar $\F$ and its conjugate $\bar \F$, 
with action
\bea
S_{\rm PM} = \int \rd^4 x \rd^2 \q  \rd^2 \bar{\q} \, \bar \F \F
-  \Big\{f  \int \rd^4 x \rd^2 \q \, \F + {\rm c.c.} \Big\}~, \qquad 
\bar D_\ad \F =0~,
\label{1.1}
\eea
where $f$ is a non-zero  parameter of mass dimension $+2$. 
This theory possesses a dual formulation described in \cite{KTyler}.
Specifically, associated with \eqref{1.1} is the first-order model  
\begin{subequations}\label{1.2}
\bea
S^{(\S)}_{\text{first-order}}&=& \int \rd^4 x \rd^2 \q  \rd^2 \bar{\q} \, \Big( \bar U U -\S U - \bar \S \bar U\Big)~, \label{1.2a} \\
-\frac 14 \bar D^2 \S &=&f~,
\label{1.2b}
\eea
\end{subequations}
in which $U$ is a complex unconstrained superfield, and $\S$ is a 
{\it deformed complex linear} superfield constrained by \eqref{1.2b}. 
Varying \eqref{1.2a} with respect to $\S$ gives $U=\F$, 
and then  the action \eqref{1.2a} reduces to \eqref{1.1}. 
Therefore, the supersymmetric theories \eqref{1.1} and \eqref{1.2} are equivalent. 
On the other hand,
the auxiliary superfields $U$ and $\bar U$  can  be integrated out 
from  the action \eqref{1.2a} resulting with 
\bea
S_\S= -\int \rd^4 x \rd^2 \q  \rd^2 \bar{\q} \,  \bar \S \S~.
\label{1.5}
\eea
The Goldstino superfield model of \cite{KTyler} made use of $\S$ subject to the holomorphic constraints
\bea
\S^2 =0\, , \qquad
-\frac 14 \Sigma{\bar D}^2D_\alpha\Sigma={f}
D_\alpha\Sigma\,.
\eea

There exists a different dual formulation for \eqref{1.1}. 
Let us consider the following first-order action
\begin{subequations} \label{1.4}
\bea
S^{({\mathfrak G})}_{\text{first-order}}&=& \int \rd^4 x \rd^2 \q  \rd^2 \bar{\q} \, \Big( 
\hf V^2 - {\mathfrak G} V \Big) ~,
\label{1.4a}
 \\
-\frac 14 \bar D^2 {\mathfrak G} &=&f~,\quad \bar{\mathfrak G} ={\mathfrak G}~.
\label{1.4b}
\eea
\end{subequations}
Here $V$ is a real unconstrained superfield, and $\mathfrak G$ is a 
{\it deformed real linear} superfield constrained by \eqref{1.4b}.
The supersymmetric theories \eqref{1.1} and \eqref{1.4} are equivalent. 
Indeed, varying 
$S^{({\mathfrak G})}_{\text{first-order}}$
with respect to $\mathfrak G$ gives $V = \F + \bar \F$, 
and then the action \eqref{1.4a} reduces to \eqref{1.1}. On the other hand, we can integrate out the auxiliary superfield $V$ from $S^{({\mathfrak G})}_{\text{first-order}}$
to end up with 
\bea
S_{\mathfrak G}= -\hf \int \rd^4 x \rd^2 \q  \rd^2 \bar{\q} \,  {\mathfrak G}^2~.
\eea

It is easy to see that requiring the deformed real linear superfield
$\mathfrak G$ to obey the quadratic nilpotency condition 
\eqref{0.0} does not allow us to eliminate the scalar field, 
${\mathfrak G}|_{\q=0}$, contained in $\mathfrak G$. However, this becomes possible 
if we subject $\mathfrak G$ to the cubic nilpotency condition
\bea
{\mathfrak G}^3 =0~.
\eea
The resulting supermultiplet contains only two fields, which are the Goldstino 
and the axion, the latter being described in terms of a gauge two-form. 
In this paper we will study the properties of this supermultiplet and its generalisations, 
including its couplings to Yang-Mills supermultiplets and supergravity.

It should be pointed out that cubic nilpotency conditions 
have been discussed in the literature 
\cite{KS,Kahn:2015mla,FKT,CKL,DallAgata:2015zxp} for two $\cN=1$ superfields, 
one of  which is the nilpotent chiral scalar $X$ subject to the only constraint 
\eqref{0.0}, as proposed in  \cite{Casalbuoni,KS}. 
Cubic nilpotency conditions 
for a single $\cN=2$ Goldstino superfield 
have been proposed in \cite{KMcAT-M,DFS,Kuzenko:2017gsc}.

This paper is organised as follows. In section 2 we show how 
deformed real linear superfields \eqref{1.4b} originate within 
a framework generalising the linear-chiral duality. 
Our new nilpotent multiplet is described in section 3. 
Its couplings to a three-form multiplet, a super Yang-Mills multiplet 
and three-form supergravity  
are presented in sections 4 and 5. 
Finally, the appendix is devoted to some  generalisations 
of the duality transformations described above.


\section{A generalisation of the linear-chiral duality} \label{section2}

We start by recalling the linear-chiral duality as described in \cite{LR}.
Consider a general two-derivative model for a  self-interacting 
$\cN=1$ tensor multiplet \cite{Siegel}
\bea
S[G] = 
\int  {\rm d}^4x {\rm d}^2 \q \rd^2 \bar \q \, \cF(G) ~, \qquad 
 D^2 {G} = \bar D^2 { G} = 0~,
\label{2.1}
\eea
where $G=\bar G$ is the gauge-invariant field strength of the tensor multiplet, 
and $\cF(x) $ is a smooth function of a real variable $x$.\footnote{Following \cite{FWZ}, $G$ is called a real linear superfield.}
The choice $\cF(x) =- x^2 $ corresponds to the free tensor multiplet \cite{Siegel}, 
while another choice $\cF(x) = - x\ln x$ corresponds to the 
so-called improved tensor multiplet \cite{deWR}.

We associate with \eqref{2.1} the following first-order model 
\bea
S[K, \F, \bar \F] = \int 
{\rm d}^4x {\rm d}^2 \q \rd^2 \bar \q \, 
\Big\{ \cF(K) -  (\F + \bar \F )K \Big\} ~, 
\qquad \bar D_\ad \F=0~.
\label{2.2}
\eea
Here the dynamical variables are a real unconstrained superfield $K$, 
a chiral scalar $\F$ and its complex conjugate $\bar \F$. 
Varying $S[K, \F, \bar \F]$ with respect to the Lagrange multiplier $\F$ gives 
the equation of motion $\bar D^2 K= 0$, and hence $K=G$. 
Then the second term in the integrand \eqref{2.2}
drops out, and we are back to the  multiplet model \eqref{2.1}. 
Therefore, the theories \eqref{2.1} and \eqref{2.2} are equivalent.
On the other hand, we can vary \eqref{2.2} with respect to 
$K$ resulting in the equation of motion 
\bea
\cF'(K) = \F + \bar \F ~.
\label{2.3}
\eea
Assuming that $\cF (x)$ possesses a Legendre transform,
this equation allows us to express $K$ as a function of $\F$ and $\bar \F$, 
and then \eqref{2.2} turns into the dual action 
\bea
S_{\rm D} [ \F, \bar \F] = \int 
{\rm d}^4x {\rm d}^2 \q \rd^2 \bar \q \, 
\cF_{\rm D}(\F + \bar \F) ~,
\label{2.4}
\eea
where $\cF_{\rm D}$ is the Legendre transform of $\cF$. 
This supersymmetric nonlinear  $\s$-model is a dual formulation 
for the tensor multiplet theory \eqref{2.1}.

There exists a variant realisation of the scalar multiplet known as 
the three-form multiplet \cite{Gates}. It is obtained by replacing 
the chiral scalar $\F$ with $\c$ given by
\bea
\c = -\frac{1}{4} \bar D^2 U ~,\qquad \bar U =U~.
\label{2.5}
\eea
Now, starting from the nonlinear $\s$-model \eqref{2.4}, 
we may construct a theory of self-interacting three-form multiplet
\bea
S_\a [ \c, \bar \c] = \int 
{\rm d}^4x {\rm d}^2 \q \rd^2 \bar \q \, 
\cF_{\rm D}( \re^{\ri \a}\c +\re^{-\ri \a} \bar \c) ~,
\label{2.6}
\eea
for some parameter $\a \in \mathbb R$. 
Since the prepotential $U$ in \eqref{2.5} is real, we cannot absorb the phase factor
$\re^{\ri \a}$ into $\c$, unlike the case of $\F$. 
Let us make the same replacement, $\F \to \re^{\ri \a}\c $, in the first-order action
\eqref{2.2}, resulting with\footnote{In the case $\cF(K) \propto K^2$, the first-order action 
\eqref{2.7} was considered in \cite{BK88}.}
\bea
S_\a[K, \c, \bar \c] = \int 
{\rm d}^4x {\rm d}^2 \q \rd^2 \bar \q \, 
\Big\{ \cF(K) -  ( \re^{\ri \a}\c +\re^{-\ri \a} \bar \c )K \Big\} ~.
\label{2.7}
\eea
This theory is equivalent to \eqref{2.6}. However, varying the action \eqref{2.7}
with respect to $U$ gives the equation
\bea
\re^{\ri \a} \bar D^2 K  +\re^{-\ri \a} D^2 K =0~,
\eea
which is  equivalent to 
\bea
 \bar D^2 K  = \ri \re^{-\ri \a} m ~, \qquad m = \bar m = {\rm const}~,
\label{2.9}
 \eea
 for some real parameter $m$. This equation defines a deformed 
real  linear multiplet. 


\section{New nilpotent multiplet}

We consider a real scalar superfield ${\bm G} = \bar {\bm G}$
subject to a deformed linear constraint
\bea
-\frac 14 D^2 {\bm G} = \bar \m = {\rm const} \quad \Longleftrightarrow \quad
-\frac 14 \bar D^2 {\bm G} = \m = {\rm const}~,
\label{3.1}
\eea
for some non-zero complex parameter $\m$. 
The general solution to this constraint is 
\bea
{\bm G} &=& \vf +\q^\a \j_\a +\bar \q_\ad \bar \j^\ad +\q^2 \bar \m +\bar \q^2  \m
+\q\s^a \bar \q H_a  
+\frac{\ri}{2}   \q^2 \pa_a  \j \s^a \bar \q
-\frac{\ri}{2}  \bar \q^2 \q \s^a\pa_a \bar \j 
\non \\
&& -\frac 14 \q^2 \bar \q^2 \Box \vf~,
\label{3.2}
\eea
where $H^a$ is the Hodge-dual of the field strength for a gauge two-form,
\bea
\pa_a H^a =0~.
\label{3.3}
\eea
From the superfield action 
\bea
S = -\int  \rd^4x \rd^2 \q  \rd^2 \bar \q\, {\bm G}^2 
\eea
we read off the component Lagrangian
\bea
\cL = - 2|\m|^2 -\hf \pa^a \vf \pa_a \vf - \ri \j \s^a \pa_a \bar \j +\hf H^a H_a~.
\label{Lagrangian} 
\eea
The constant term in $\cL$ indicates that a positive cosmological constant is generated 
once the system is lifted to supergravity.

We also subject $\bm G$ to 
the cubic nilpotency condition
\bea
{\bm G}^3 =0~.
\label{3.4}
\eea
The top component of this constraint can be written in the form 
\begin{subequations}\label{3.5}
\bea
(a - \frac 14 \vf \Box \vf ) \vf = \hf b~,
\label{3.5a}
\eea
where 
\bea
a&=& 2|\m |^2 +\frac{\ri}{2} ( \j \s^a \pa_a \bar \j -  \pa_a \j \s^a \bar \j ) 
- \hf H^aH_a~,\label{3.5b}\\
b &=&  \m \j^2 + \bar \m \bar \j^2 - H^a \j \s_a \bar \j~.
\label{3.5c}
\eea
\end{subequations}
It holds that 
\bea
b^2 = a \j^2 \bar \j^2 ~, \qquad b^3 =0~.
\eea
Equation \eqref{3.5a} analogous to the one derived in \cite{DFS} 
for the case of $\cN=2 \to \cN=1$ spontaneous supersymmetry breaking.
Therefore, eq. \eqref{3.5a} may be solved similarly to the approach employed in \cite{DFS}. Specifically, we have to look for a solution of the form 
\bea
 \vf = \cU \j^2 + \bar \cU \bar \j^2 + \cV^a \j \s_a\bar \j~,
\label{3.7}
 \eea
 with $\cU$ and $\cV^a$ being some composites of the dynamical fields $\j_\a$, 
 $\bar \j_\ad$ and $H^a$. The point is that it is the ansatz \eqref{3.7} which is 
 consistent with the two lowest components of \eqref{3.4}, which are $\vf^3 =0$, 
  $\vf^2 \j_\a =0$ and $\vf^2 \bar \j_\ad =0$. Now it follows from \eqref{3.5a} that 
 \bea
 \vf^2 = \frac{\j^2 \bar \j^2}{4a}= \Big(\frac{b}{2a}\Big)^2~.
 \eea
 This relation implies that the general solution of \eqref{3.5a} is 
 \bea
 \vf = \frac{b}{2a} +  \frac{b^2}{32a^4} \Box b~.
\label{3.11}
 \eea
The solution is well defined provided 
\bea
a_0:= 2|\m |^2- \hf H^aH_a \neq 0~.
\eea

Making use of  \eqref{3.11}, 
the Lagrangian \eqref{Lagrangian} turns into 
\bea
\cL = - 2|\m|^2  +  \frac{b}{8a} \Box  \frac{b}{a}
+ \frac{b^2}{64 a^5} (\Box b )^2
- \ri \j \s^a \pa_a \bar \j +\hf H^a H_a~,
\label{Lagrangian2} 
\eea
modulo a total derivative. This Lagrangian depends on $H^a$ in a highly 
nonlinear way. However,  all nonlinear contributions contain fermionic factors 
of $\j_\a$ and $\bar \j_\ad$. As a result, it is possible to dualise the gauge 
two-form, described by its gauge-invariant field strength $H^a$,
 into an axion $\cS$ by considering the first-order model
\bea
\cL_{\text{first-order}} = \cL (H) - H^a \pa_a \cS ~,
\eea
in which $H^a$ is an unconstrained vector field, and $\cL (H)$
stands for the Lagrangian \eqref{Lagrangian2}.


\section{Nilpotent tensor multiplet coupled to a three-form multiplet}

The construction given in the pervious section admits a natural generalisation. 
The idea is that the complex parameter $\m $ in \eqref{3.1} may be viewed as 
the expectation value of a chiral superfield. Therefore, a more general constraint
is given by 
\bea
-\frac 14 \bar D^2 {\bm G} = \cY~, \qquad \bar \cD_\ad \cY =0~,
\label{4.1}
\eea
for some background chiral superfield 
\bea
\cY (x,\q,\bar \q) = \re^{\ri \q \s^a \bar \q \pa_a} \Big( \m (x) + \q^\a \r_\a(x) 
+ \q^2 F(x)\Big)~.
\label{4.2}
\eea
Due to the identity 
\bea
-\frac 14 \big[ D^2 , \bar D^2 \big] = \ri \pa_{\a\ad}\big[D^\a  , \bar D^\ad \big] ~,
\eea
in order for ${\bm G} $ to contain a conserved vector field, 
 $\cY$ has to be  a three-form multiplet, which means that locally 
\bea
\cY=-\frac{1}{4}\DB^2 U~,~~~~~~
\bar{U}= U~,
\label{4.4}
\eea
where $U$ is a real but otherwise unconstrained prepotential.
The only implication of the representation 
\eqref{4.4} is that the auxiliary field $F$ in \eqref{4.2} 
is not an arbitrary complex field, but instead has the form 
\bea
F = d+\frac{\ri}{2} \pa_a c^a~,
\eea
where $d$ is a real scalar, and $c^a$ is the Hodge-dual of a gauge three-form. 
The general solution to the constraint \eqref{4.1} is 
\bea
{\bm G} &=& \vf +\q^\a \j_\a +\bar \q_\ad \bar \j^\ad +\q^2 \bar \m +\bar \q^2  \m
+\q\s^a \bar \q \big(H_a +c_a\big)  \non \\
&&  + \q^2 \big( \bar \r   +\frac{\ri}{2}\pa_a  \j \s^a \big)\bar \q
+ \bar \q^2 \q \big(\r  -\frac{\ri}{2}  \s^a\pa_a \bar \j \big)
+ \q^2 \bar \q^2 \big( d-\frac 14 \Box \vf \big)~,
\eea
where $H_a$ obeys the constraint \eqref{3.3}.

As in the previous section, we impose the cubic nilpotency condition
\bea
{\bm G}^3 =0~.
\label{4.7}
\eea
The top component of this constraint can be written in the form 
\begin{subequations}
\label{4.8}
\bea
\Big( \hat a + \vf \big(d - \frac 14  \Box \vf \big) \Big)  \vf = \hf \hat b~,
\label{4.8a}
\eea
where 
\bea
\hat a&=& 2|\m |^2 -  \j  \big( \r -\frac{\ri}{2}\s^a \pa_a \bar \j \big)
-\big( \bar \r+  \frac{\ri}{2}\pa_a \j \s^a \big)\bar \j  
- \hf (H+c)^2~,
\label{4.8b}
\\
\hat b &=&  \m \j^2 + \bar \m \bar \j^2 - \big(H^a +c^a\big)\j \s_a \bar \j~.
\label{4.8c}
\eea
\end{subequations}
As in the previous section,  the lowest components of \eqref{4.7},  
$\vf^3 =0$,   $\vf^2 \j_\a =0$ and $\vf^2 \bar \j_\ad =0$, 
imply that $\vf$ has to have the form \eqref{3.7}
Now it follows from \eqref{4.8a} that 
 \bea
 \vf^2 = \frac{\j^2 \bar \j^2}{4\hat a}= \Big(\frac{\hat b}{2\hat a}\Big)^2~.
 \eea
 This relation implies that the general solution of \eqref{3.5a} is 
 \bea
 \vf = \frac{\hat b}{2\hat a} - \frac{\hat b^2}{4\hat a^3} d
 +  \frac{\hat b^2}{32\hat a^4} \Box \hat b~.
 \eea
 
 As an example of the construction given, we can choose $\c$ of the form 
\bea
\c = \m + g\, \tr (W^\a W_\a) ~, 
\eea
where $g$ is a  real parameter, and 
 $W_\a$ the covariantly chiral field strength of a Yang-Mills supermultiplet.
 Essentially, we are in a position to recycle the classic results on Cher-Simons 
 couplings for linear multiplets, see, e.g., \cite{BGG} for a review.  
 

\section{Coupling to three-form supergravity} 

As is well known, every off-shell  formulation for $\cN=1$ supergravity can be realised as $\cN=1$ conformal supergravity coupled to a compensating multiplet
(see, e.g., \cite{GGRS,Ideas} for reviews).
Different off-shell formulations correspond to choosing different 
compensators. 
As reviewed in \cite{Ideas},  
conformal supergravity can be described 
using the superspace geometry of \cite{GWZ}, 
which underlies the Wess-Zumino approach \cite{WZ,Zumino}
to old minimal supergravity
\cite{old1,old2}.
This requires to extend the supergravity gauge group to include the super-Weyl 
transformations introduced in \cite{HT}. For the technical details, we refer the reader to 
the textbook \cite{Ideas}, see also the recent paper \cite{KT-M17-2}. 
The notation and conventions of \cite{Ideas} are used throughout this paper.

We start by reviewing the super-Weyl invariant formulation for
three-form supergravity \cite{GS,OvrutWaldram}, which was 
given in \cite{KMcC}.\footnote{This formulation has been used in 
recent publications \cite{BK17,KT-M17-2,FLMS}.} 
The corresponding  conformal compensator is a  three-form multiplet
coupled to conformal supergravity.
It is described by a covariantly chiral scalar $\P$ and its conjugate $\bar \P$, 
with $\P$ defined by  
\bea
\P= -\frac{1}{4}\big(\cDB^2-4R\big) P~,
\qquad {\bar P} =P
~,
\label{F4-P}
\eea
where 
the scalar prepotential $P$ in \eqref{F4-P}
is real but otherwise unconstrained.\footnote{The operator 
 $\bar{\D}:=-\frac{1}{4}\big(\cDB^2-4R\big)$ is the covariantly chiral projection 
 operator introduced in \cite{WZ,Zumino}.}
The compensator $\P$ has to be nowhere vanishing so that $\P^{-1}$ exists.
We postulate  $P$ to be super-Weyl primary of  weight $(1,1)$, 
\begin{subequations}\label{s-weyl-P}
\bea
\d_\s P &=& (\s+\bar\s)  P~,
\eea
which implies that $\P$ is also primary, 
\bea
\d_\s \Pi &=& 3\s\Pi~.
\eea
\end{subequations}
As follows
 \eqref{F4-P}, the prepotential $P$ is defined modulo
 gauge transformations of the form 
\bea
\d_L P = L~, 
 \qquad  
\big(\cDB^2-4R\big)
L
=0~, \qquad \bar L =L
~.
\label{gauge-inv-P}
\eea
The gauge parameter $L$ is a 
covariantly real  linear superfield.

The action for three-form supergravity is 
\bea
S_{\text{SG}} &=&-\frac{3}{\k^2} \int \rd^4x\rd^2\q\rd^2\qb
\,E\,\Big\{
\big(\bar \P \P\big)^{\frac{1}{3}}
- \hf mP
\Big\} \non \\
&=& -\frac{3}{\k^2} \int \rd^4x\rd^2\q\rd^2\qb
\,E\,
\big(\bar \P \P\big)^{\frac{1}{3}}
 + \Big\{ \frac{m}{\k^2} \int {\rm d}^4x {\rm d}^2 \q \,\cE\,   \P  + {\rm c.c.} \Big\}
~,
\label{3-form_sugra}
\eea
where $m$ is a real parameter, and $E$ and $\cE$ denote the full superspace 
and the chiral subspace integration densities, respectively.
By construction the action is invariant under 
gauge transformations \eqref{gauge-inv-P}.

Complex three-form supergravity \cite{old1,GS,FLMS} is obtained 
by choosing the prepotential $P$ in \eqref{F4-P} to have the form 
\bea
P= \G + \bar \G~,
\eea
where  $\G$ is a covariantly complex linear scalar superfield constrained by 
\bea 
\big(\cDB^2-4R\big)
\G=0
~.
\label{complex-linear}
\eea
Due to this constraint, the field strength \eqref{F4-P} reads 
\bea
\P= -\frac{1}{4}\big(\cDB^2-4R\big) \bar \G~.
\eea

The general solution to the constraint 
\eqref{complex-linear} 
is known \cite{GGRS} to be 
\bea
\G = \cDB_\ad \bar{\J}^\ad~,
\eea
where $\bar \J^\ad $ is an unconstrained spinor superfield defined modulo 
gauge transformations 
\bea
\d_\L \bar \J^\ad = \bar \cD_\bd \bar \L^{(\ad \bd)}~,
\eea
which leave $\G$ invariant. The super-Weyl transformation of $\bar \J^\ad $
is chosen to be \cite{Ideas}
\bea
\d_\s \bar \J^\ad =\frac{3}{2} \bar \s \bar \J^\ad~,
\eea
and this transformation law implies 
\bea
\d_\s\G=(\s+\bar{\s})\G 
\eea

We define a deformed covariantly linear multiplet to obey the constraint 
\bea
 -\frac{1}{4}\big(\cDB^2-4R\big) {\bm G} = f \P + \c~, \qquad 
 \bar \cD_\ad \c=0~, \qquad 
 f = {\rm const}~.
 \label{5.12}
\eea
Here $\c$ is a super-Weyl primary three-form multiplet, which means 
the following: (i) 
the super-Weyl transformation of $\c$ is 
\begin{subequations}
\bea
\d_\s \c &=& 3\s \c~;
\eea
and (ii) $\c$ has the property 
\bea
{\rm Im}  \int {\rm d}^4x {\rm d}^2 \q \,\cE\,  \c =0~.
\eea
\end{subequations}
For instance, we can choose $\c$ of the form 
\bea
\c = g_1\, \tr (W^\a W_\a) +g_2 \,W^{\a\b\g } W_{\a\b\g}~, 
\eea
where $g_1$ and $g_2$ are real parameters, and 
 $W_\a$ the covariantly chiral field strength of a super Yang-Mills multiplet, 
 and $W_{\a\b\g}$ is the super-Weyl tensor \cite{GWZ}, see \cite{Ideas} for more details.
The non-zero parameter $f$ in \eqref{5.12} is real (complex) provided
the three-form multiplet $\P$ is real (complex). 
As in the rigid supersymmetric case, we subject $\bm G$ to the nilpotency condition
\bea
{\bm G}^3 =0~.
\eea
The action for the nilpotent tensor multiplet  is 
\bea
S_{\text{AG}} &=&- \int \rd^4x\rd^2\q\rd^2\qb
\,E\,{{\bm G}^2}{\big(\bar \P \P \big)^{-\frac{1}{3} }}
\eea
The complete supergravity-matter action is
$S= S_{\rm SG} + S_{\text{AG}} +S_{\rm SYM}$, where $S_{\rm SYM}$
denotes the standard super Yang-Mills action in the presence of supergravity \cite{WZ}.
\\

\noindent
{\bf Acknowledgements:}\\
The author is grateful to Gabriele Tartaglino-Mazzucchelli for comments 
on the manuscript. Discussions with
Dmitri Sorokin and Tsutomu Yanagida are gratefully acknowledged.
The research presented in this  work is supported in part by the Australian 
Research Council, project No. DP160103633.


\appendix 

\section{More duality transformations}

In this appendix we  discuss some  generalisations 
of the duality transformations described in section \ref{section1}. 
As an extension of the chiral model \eqref{1.1}, we consider
\bea
S = \int \rd^4 x \rd^2 \q  \rd^2 \bar{\q} \, \bar \F \F
-  \Big\{  \int \rd^4 x \rd^2 \q \, \J \F + {\rm c.c.} \Big\}~, \qquad 
\bar D_\ad \c =0~,
\label{A.1}
\eea
where $\J$ is a background chiral superfield. This theory has a dual formulation 
 that can be  obtained by making use of the first-order action \eqref{1.2a}, in which
 $\S$ has to obey the constraint\footnote{Constraints  
of the form (\ref{A.2}) were introduced for the first time by Deo and Gates \cite{DG85}.
In the context of supergravity, such constraints 
were used in \cite{K15,KTyler} to generate
couplings of the complex linear Goldstino superfield to chiral  matter.}
 \bea
 -\frac 14 \bar D^2 \S = \J~,
 \label{A.2}
 \eea
which is a deformation of \eqref{1.2b}. 
The dual action has the form  \eqref{1.5} with $\S$ 
constrained according to  \eqref{A.2}. 
There exist dual formulations for chiral models that 
are obtained from \eqref{A.1} by replacing 
the superpotential by the rule $\J \F \to \J \F^n$, 
for an integer  $n>2$.  The dual actions are described in \cite{BK11,KT-M11}
(such actions requires $\F$ to be nowhere vanishing).

A different duality transformation exists if the
background chiral scalar $\J$ in \eqref{A.1} is a three-form multiplet,
\bea
\J =-\frac 14 \bar D^2 U ~, \qquad \bar U =U~.
\label{A.3}
\eea
Then the action \eqref{A.1} can be rewritten as an integral
over the full superspace, 
\bea
S = \int \rd^4 x \rd^2 \q  \rd^2 \bar{\q} \, \Big\{ \bar \F \F
-U (\F +\bar \F ) \Big\}~.
\label{A.4}
\eea
This action is obviously invariant under gauge transformations
of the prepotential $U$ of the form 
\bea
\d_L U = L ~, \qquad \bar D^2 L =0~,
\label{A.5}
\eea
with the chiral scalar $\J$ defined by \eqref{A.3} being a gauge-invariant field strength. 
Since the action \eqref{A.4} depends on $\F$ and $\bar \F$ only via the combination
$\F + \bar \F$, the model naturally possesses a dual formulation given in terms of a real linear superfield $G$, see section \ref{section2}. The dual action is 
\bea
S = -\hf \int \rd^4 x \rd^2 \q  \rd^2 \bar{\q} \, (G+U)^2~.
\label{A.6}
\eea
It is invariant under the gauge transformation \eqref{A.5} provided $G$
transforms as
\bea
\d_L G = -L~.
\eea
The action \eqref{A.6} is constructed  in terms of the gauge-invariant 
superfield  ${\frak G} := G +U =\bar{\frak G}$
obeying the constraints
\bea
 -\frac 14 \bar D^2 {\frak G} = \J~,
 \label{A.8}
 \eea
which is a deformation of \eqref{1.4b}.
The same model can  be naturally obtained by considering the first-order action 
\eqref{1.4a} in which $\frak G $ is subject to the constraint \eqref{A.8}.


\begin{footnotesize}

\end{footnotesize}

\end{document}